\documentclass[12pt]{article}
\usepackage{a4}
\usepackage{array}
\usepackage{fancybox}
\usepackage{hhline}
\usepackage{amsmath,amssymb}
\usepackage{graphics}  
\def\rc{\hbox{$\mathcal C $}}

\def\rf{\hbox{$\mathcal F $}}

\def\rk{\hbox{$\mathcal K $}}
\def\rp{\hbox{$\mathcal P $}}
\def\rm{\hbox{$\mathcal M $}}
\def\rq{\hbox{$\mathcal Q $}}

\def\rs{\hbox{$\mathcal S $}}
\def\tr{\hbox{$\mathcal T $}}

\def\nf{\hbox{$\mathcal{ \cal N}_f$}}

\begin{document}
\title{Generalisation of DGLAP equations to massive partons}
\author{C. Pascaud\\
LAL,Universit\'e Paris-Sud,CNRS/IN2P3,Orsay,France\\
pascaud@lal.in2p3.fr}
\maketitle
\begin{abstract}
DGLAP evolution equations are modified in order to use all the quark families in the
full scale range, satisfying kinematical constraints and sumrules, thus
having complete continuity for the pdfs and observables. Some consequences
of this new approach are shown.

\\ 

{\bf Comments:} 12 Pages and 5 Figures
\end{abstract}
\section{Introduction} 
 As it is well known heavy quarks present a challenge in the phenomological
description of deep inelastic scattering of leptons against nucleons:
DGLAP \cite{a-p} evolution
equations are essential ingredients to this description but they consider
only massless partons.
With the advent of HERA results, the increase of the statistical
precision of the measurements and the prespectives open by LHC,
the necessity to have a way to predict
the observables from the lowest to the highest $Q^2$ scale became clear
and several so-called variable flavor number schemes (VFNS) have
appeared:
\begin{itemize}
\item
In the massless approach a flavor $h$ is usually considered to be active when
its quark mass $m^2_h$ is smaller than the scale $Q^2$.
Up to this limit its parton distributions
(for quark and antiquark) are null, after the
number of flavors is increased by one and they start to evolve from zero.
\item
 In the massive approach the number of flavors is considered to be constant
but  the heavy quark  is produced by the existing partons \
thus it have structure functions coming from $Q^2$ dependent coefficient
functions and no (pdf) distributions.
\item
VFNS  approaches make use usually of the two preceeding approaches according
to predefined $Q^2$ regions: At some arbitrary fixed value
the number of quark flavors is increased by one and the pdfs are rearranged 
according to theory inspired rules before resuming evolution.
All this is a still an ongoing activity.
\end{itemize}
The path of this work is different: It will consider always
six quark species but modify coherently DGLAP equations, splitting functions,
coefficient functions,flavor
number and running $\alpha_s$ in order to have all the parton distributions
and structure functions continous in the full kinematical range.
This will be done furthermore in such a way that when the
phase space  for an heavy quark is quasi null ($ m^2_h \gg Q^2$ )
its distributions are quasi null and its influence on the full system
is also quasi null. On the contrary when its phase space is quasi complete
 ($ m^2_h \ll Q^2$ ) it will behave like a light quark. The key to this will be
kinematic constraints and sumrules.\\
To fulfill these commitments, section 2  presents
DGLAP equations in an appropriate manner,
enphasizing  the components which will need modification.
Section 3 will present the modified (extended) equations defining the method
thereafter called cfns (continous flavor number scheme),
section 4 brings arguments to justify the method, 
 section 5 shows a possible program implementation
and finally Section 6 will present a comparaison 
between massless scheme and cfns at NNLO in $\alpha_s$.

\section {Kinematics and DGLAP equations}
The electron proton reaction is:
$$e(l) + P(p) \rightarrow e(l-q)+X(p+q)$$
Within the symbols ( ) are quadrimomenta.\\
Kinematic variables are:\\
$S=(l+p)^2$ \hspace{5mm}$Q^2 = Sxy = -q^2>0 $ 
\hspace{5mm} $x=\frac{Q^2}{2p.q}$ \hspace{5mm}$y=\frac{p.q}{p.l}$ \hspace{5mm}
$W^2=(p+q)^2=M^2+Q^2(\frac{1}{x}-1)$\\
$M$ and $W$ are the initial and final  hadronic masses.\\
Parton $o$ may be kicked out of the target if the final hadronic mass
is $W>2m_o$ where $ m_o$ is the parton $o$ mass. 
This translate into a kinematic limit $x<l_o$ with 
$l_o=\big(1+4m_0^2Q^{-2}\big)^{-1}$\\
Light partons fulfill always this condition
but heavy quarks do only for $Q^2 \rightarrow \infty$.\\
Usually the flavor number $\nf$ is taken as the number of quark families
such that $Q^2>m_o^2$ : $\nf=N_f$ integer.
$\nf$ is the main concern of the overall approach.\\
 DGLAP equations read~:
\begin{equation}\label{dglap}
\frac{\partial o(Q^2)}{\partial ln(Q^2)}
=  \sum_i \rp_{oi} \otimes i(Q^2)
\end{equation}
$\rp$ are splitting functions, $i$ and $o$ are parton distributions (pdf)
 and run on the $1+2N_f$ partons species, they will be noted also
by the name of their species $p=$ $g$,$d$,$\bar d$,$u$,$\bar u$,...
and for the quarks $d^\pm=d\pm\bar d$,... will be introduced.\\
All these are $x$ and $Q^2$ functions.
$\otimes$ note the convolution between two functions of $x$ defined by:
$$[A \otimes B ] (x) = \int_x^1 A(z)B(\frac{x}{z})\frac{dz}{z}
= \int_x^1 A(\frac{x}{t})B(t)\frac{dt}{t}
$$
%Notice that integration limits may be extended to $0,\infty$ if the
%functions argument $\in 0,1$\\
The following properties hold:\\
$A \otimes B =B \otimes A   \qquad 
[A \otimes B] \otimes C=A \otimes [B \otimes C]  \qquad 
x^n [A \otimes B] = [x^n A] \otimes [x^n B]$\\
$\int_0^1 [A \otimes B] dx = \int_0^1 A(x) dx \int_0^1 B(x) dx$\\
$\int_0^1 \delta(x-l)dx = 1 \qquad \int_0^1 x \delta(x-l)dx = l \qquad 
\delta(x-l)\otimes A = \frac{1}{l} A( \frac{x}{l}) $
%\qquad $\delta(x-l_1)\otimes \delta(x-l_2) = \delta(x-l_1 l_2)$
\\
\\
DGLAP equations  separate into two independent subsystems the first one being:
 \begin{eqnarray}
 \frac{\partial g (Q^2)}{\partial ln(Q^2)}=
 \rp_{gg} \otimes g (Q^2)
+\sum_{q=d}^{t} \rp_{gq} \otimes q^+ (Q^2)\\ 
 \frac{\partial q^+ (Q^2)}{\partial ln(Q^2)}=
 \tilde{\rp}_{qg} \otimes g (Q^2)
+\rp_{NS}^+ \otimes q^+ (Q^2)
+\sum_{r=d}^{t} \tilde{\rp}_S^+ \otimes r^+ (Q^2)
\end{eqnarray}
Using definitions inspired by \cite{Neerven}:
$\rp^{\pm}_{NS}=\rp^V_{qq} \pm \rp^V_{q{\bar q}}$,
$\rp^{\pm}_S=\rp^S_{qq} \pm \rp^S_{q{\bar q}}$.
$\tilde{\rp}_{qg}=\frac{\rp_{qg}}{\nf}$,
$\tilde{\rp}^{\pm}_S=\frac{\rp^{\pm}_S}{\nf}$.\\
The second system may be deduced from the first by replacing superscript
$+$ by $-$ and suppressing
all references to gluon $g$. So it  will not be mentioned anymore
and even superscript $+$ will not be written in the following.\\
Kernels $\rp$ are polynomials in $a_s=\frac{\alpha_s}{4\pi}$ and $\nf$
 as follows:\\
\footnotesize
\hspace{-15mm}
\begin{tabular}{l l l l l l }
$\rp_{gg}
=a_s   (\rp_{gg}^{00}     +\nf\rp_{gg}^{01})
$&$+a_s^2 (\rp_{gg}^{10}  $&$+\nf\rp_{gg}^{11})
$&$+a_s^3 (\rp_{gg}^{20}  $&$+\nf\rp_{gg}^{21}    $&$+\nf^2 \rp_{gg}^{22}~)$
\\
$\rp_{qg}
=a_s                       \nf\rp_{qg}^{01}
$&$+a_s^2                 $&$ \nf\rp_{qg}^{11} 
$&$+a_s^3                 $&$ (\nf\rp_{qg}^{21}    $&$+\nf^2 \rp_{qg}^{22}~)$
\\
$\rp_{gq}
=a_s   \rp_{gq}^{00}                                                 
$&$+a_s^2 (\rp_{gq}^{10}  $&$+\nf\rp_{gq}^{11})
$&$+a_s^3 (\rp_{gq}^{20}  $&$+\nf\rp_{gq}^{21}    $&$+\nf^2 \rp_{gq}^{22}~)$
\\
$\rp_{qq}^V
=a_s   \rp_{qq}^{V00}    
$&$+a_s^2 (\rp_{qq}^{V10} $&$+\nf\rp_{qq}^{V11})
$&$+a_s^3(\rp_{qq}^{V20}  $&$+\nf\rp_{qq}^{V21}   $&$+\nf^2 \rp_{qq}^{V22})$
\\
$\rp_{q\bar q}^V
=  
$&$a_s^2 (\rp_{q\bar q}^{V10} $&$+\nf\rp_{q\bar q}^{V11})
$&$+a_s^3(\rp_{q\bar q}^{V20}  $&$+\nf\rp_{q\bar q}^{V21}   $&$+\nf^2 \rp_{q\bar q}^{V22})$
\\
$\rp_S^+
= 
$&$a_s^2 (                     $&$ \nf\rp_S^{+11})
$&$+a_s^3(                     $&$\nf\rp_S^{+21}   $&$+\nf^2 \rp_S^{+22})$
\\
$\rp_S^-
=                              
$&$                            $&$
$&$a_s^3(              $&$\nf\rp_S^{-21}   $&$+\nf^2 \rp_S^{-22})$
\end{tabular}\\
\normalsize
Notice that $\rp_{gg}^{01}$ existence comes from the $\beta_0$  term in
  $\rp_{gg}$.
\section {Modified DGLAP equations}
\begin{figure}[here]
 \resizebox{1.1\textwidth}{.7\textwidth}
           {\includegraphics{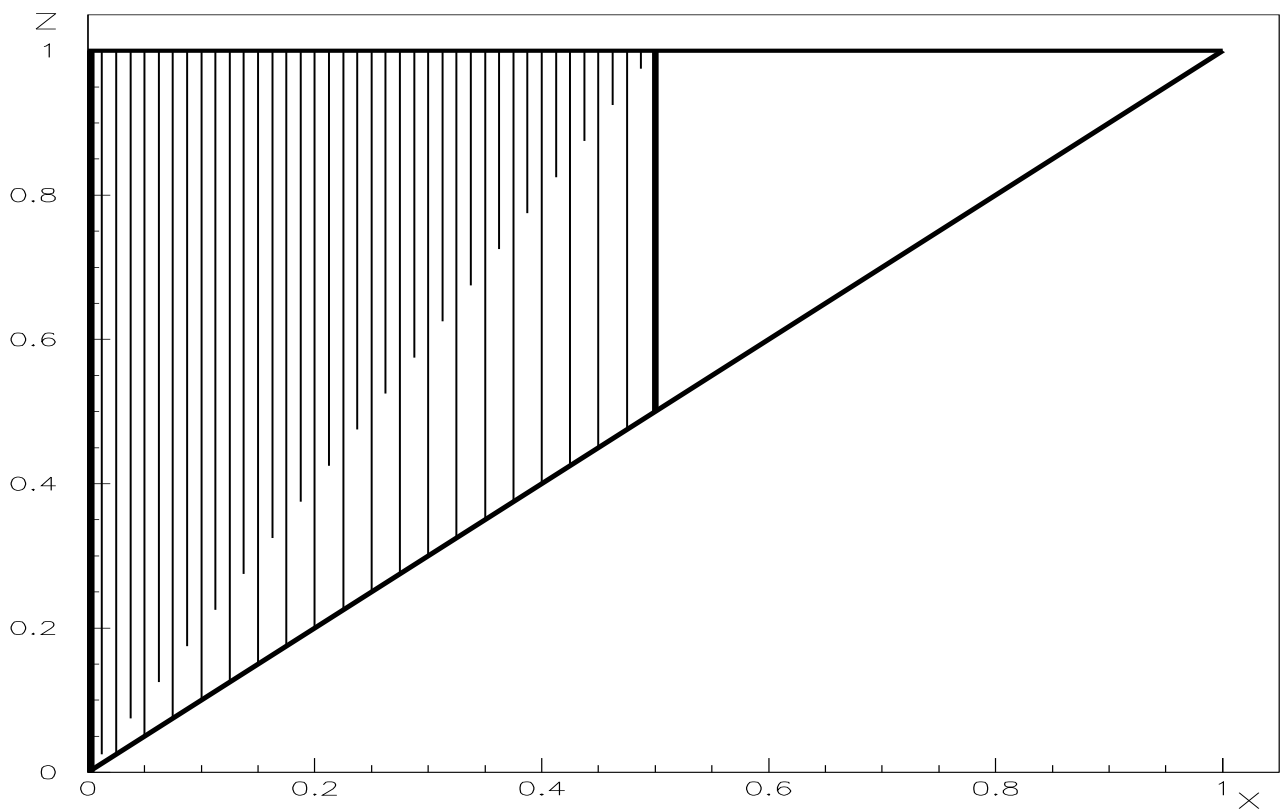}}
\caption{$C \rightarrow B $ a problematic changing term. Convolution domain
of  Equation \ref{domain}. Vertical thin spaced lines show the kinematically
reduced domain.}
\label{kine}
\end{figure}

The idea is to modify the kernels, keeping them functions
of a single argument (apart from $Q^2$ dependance coming from 
$\nf$ and $a_s$) in order
to satisfy simultanously the three kinematical constraints
$x_o<l_o$, $x_i<l_i$, $x_o<x_i$. \\
$\rp_{oi}$ gives the change to outcoming parton $o$ at Bjorken $x$ radiated by
incoming parton $i$ at Bjorken $\frac{x}{z}$.\\
Problematic cases are when parton $o$ is heavier than parton $i$ like for
$c \rightarrow b$ as shown Figure \ref{kine} where
 the corresponding term in equation(\ref{dglap}) is depicted. For a given value
of $x$ the convolution integral variable $z$ runs vertically inside the
big triangle at least for a massless parton. But for a massive quark the 
rightmost triangle has to be removed and if one wants to keep the splitting
nature of the convolution only the leftmost hached triangle has to be kept.
Notice that this will also supress an unwanted discontinuity of
that term  at $x=l_o$ and bring it gently to 0.\\
An other way to present the modification is to
replace $\rp$ by $\rs$ in the problematic changing term:
\begin{equation}\label{domain}
\int^1_x \rs_{oi}(\frac{x}{t})i(t)\frac{dt}{t}
\end{equation}
Requesting this term to be null for  $x \ge l_0$ means that
$\rs(u)=0$ for $u \ge l_o$
which is satisfied by:
\begin{equation}
\rs_{oi}=\rp_{oi} \otimes \rk_{oi} 
\end{equation}
With the definition $\rk_{oi}=f_{oi}\delta(x-l_o)$, where $f_{oi}(Q^2)$ 
is not decreasing and goes to 1   for $Q^2 \rightarrow \infty$\\
Notice that the effect of the $\delta$ function is to replace
$\rp(x)$ by $\rp(\xi)$ with $x=\xi l_o$\\
With this modification the  subsystem becomes:\\
 \begin{eqnarray}
 \frac{\partial g (Q^2)}{\partial ln(Q^2)}=
 \rp_{gg} \otimes \rk_{gg} \otimes g (Q^2)
+\sum_{q=d}^{t} \rp_{gq} \otimes \rk_{gq} \otimes q (Q^2)\\
 \frac{\partial q (Q^2)}{\partial ln(Q^2)}=
 \tilde{\rp}_{qg} \otimes \rk_{qg} \otimes g (Q^2)
+\rp_{NS} \otimes \rk_{qq}^{NS} \otimes q (Q^2)
+\sum_{r=d}^{t} \tilde{\rp}_S \otimes \rk_{qr} \otimes r (Q^2) 
\end{eqnarray}
%Notice that those equations may be more general:
%if the mass dependent kernel $\rs$ is known one may write ($\rp_{oi}$ being
%invertible)
%\begin{equation}
%\rk_{oi}=(\rp_{oi})^{-1} \otimes \rs_{oi} 
%\end{equation}
\subsection{Sumrule constraints}
In the above equations $\rp$'s are functions of the $\nf$ to be defined now.
The momentum sumrule imposes constraints which have to be satisfied (and are
for the standard DGLAP). They are obtained by requesting that the first
momentum of the sum of all parton distributions is constant and equal to 1
for any $Q^2$ value. Introducing 
$\Delta=\int_0^1 \rk xdx$ and $\rq=\int_0^1 \rp xdx $,
taking the $Q^2$ derivative of this momentum, using
the DGLAP equations and the property that the n-momentum of a convolution
product is equal to the product of the n-momenta of its components,
one find easily:
\begin{eqnarray}
\rq_{gg} \Delta_{gg} +  \frac{\rq_{qg}}{\nf} \sum_{q=d}^{t} \Delta_{qg} =0\\
\rq_{gq} \Delta_{gq} +\rq_{NS}\Delta_{qq}^{NS}
+ \frac{\rq_S}{\nf} \sum_{r=d}^{t} \Delta_{rq} =0
\end{eqnarray}
Using the fact that those equations are satisfied by the original DGLAP
equations where all the $\Delta$s are 1, one gets:
\begin{equation}
\nf=\frac{\sum_{q=d}^{t} \Delta_{qg}}{\Delta_{gg}}
= \frac{\rq_{gq}+\rq_{NS}}{\rq_{gq} \Delta_{gq} +\rq_{NS}\Delta_{qq}^{NS}}
\sum_{r=d}^{t}\Delta_{rq}
 \end{equation}
Leading to the final extended DGLAP equations \footnotemark~
\footnotetext{\it The simplest hypothesis  has been used: $f_{io}=~
\Delta_{gg}=~\Delta_{qq}^{NS}=~1$.}
\begin{eqnarray}
 \frac{\partial g(x,Q^2)}{\partial ln(Q^2)}=
 \rp_{gg} \otimes g(x,Q^2)
+\sum_{q=d}^{t} \rp_{gq} \otimes q^+(x,Q^2)\\ 
 \frac{\partial q^+(x,Q^2)}{\partial ln(Q^2)}=
 \rk_q \otimes \bigl[ \tilde{\rp}_{qg} \otimes g(x,Q^2)
+\sum_{r=d}^{t} \tilde{\rp}_S^+ \otimes r^+(x,Q^2) \bigr]
+\rp_{NS}^+ \otimes q^+(x,Q^2)
\end{eqnarray}
Where $\rk_q=\delta(x-l_q)$~.
Furthermore 
$\rq(\rp)$ being polynomials in the flavor number, the only solution
is to redefine the latter as being $\nf = \sum_{q=d}^{t} l_q$.\\
Quark kinematical limit or longitudinal phase space $l_q$ then may be also
viewed as quark activity going smoothly  from 0 to 1.
\subsection {Subsystem decoupling}
As it is well known the usual DGLAP equations decouple in
a $g$,$\Sigma$ system of two coupled equations and several non-singlet
independent equations. It is now not so simple. The  $g$,$\Sigma$ singlet
system still decouple but
%only due to the fact that in the quark equations term
%$ \rk_q \otimes  \tilde{\rp}_S^+ \otimes q^+(x,Q^2) $ has been kept altough
%it was not requested by kinematical constraint and could have been replaced by
%$$ l_q \tilde{\rp}_S^+ \otimes q^+(x,Q^2)$ and second 
the non-singlets cannot all
decouple. In fact there is not anymore a unique way to simplify the system.
The following solution is one of the both
which have been realised in the programn implementation section 5.\\
Define:\\
$\rk=\sum_{q=d}^t \rk_q$
\hspace{20mm}
$\Sigma_L = \sum_{q=d}^s q $
\hspace{20mm}
$l_{LN} = l - \frac{\Sigma_L}{3}$\\
Subscript $_{LN}$ is used do distinguish these non singlets from the usual
ones which may not be used here with $\nf$ varying continously.
With these one get the following subsystem:
 \begin{eqnarray}
 \frac{\partial g (Q^2)}{\partial ln(Q^2)}&=&
 \rp_{gg} \otimes  g (Q^2)
+ \rp_{gq} \otimes  \Sigma (Q^2)\\
   \frac{\partial  \Sigma (Q^2)}{\partial ln(Q^2)}&=&
\rk \otimes\tilde{\rp}_{qg} \otimes  g (Q^2)
+  \bigl( \rp_{NS}+\rk \otimes \tilde{\rp_S} \bigr) \otimes \Sigma (Q^2)\\
  \frac{\partial  l_{LN} (Q^2)}{\partial ln(Q^2)}&=&
                                 \rp_{NS} \otimes  l_{LN} (Q^2)\\
   \frac{\partial h  (Q^2)}{\partial ln(Q^2)}&=&
 \rk_h \otimes\tilde{\rp}_{qg} \otimes  g (Q^2)
+                                 \rp_{NS} \otimes  h (Q^2)
+\rk_h \otimes \tilde{\rp_S}  \otimes \Sigma (Q^2)
\end{eqnarray}
After evolution of the seven pdfs  $g$,$\Sigma$,$c$,$b$,$t$,$d_{LN}$,$u_{LN}$
the full system may be recovered using:\\
$\Sigma_L=\Sigma-c-b-t$ \hspace{30mm}
$d_{LN}+u_{LN}+s_{LN}=0$\\
\vspace{1mm}
Note that for $l_h \rightarrow 0$ the corresponding DGLAP equation will get
decoupled (Appelquist-Carazzone theorem) and the kinematical constraint automatically verified.\\

\subsection{Renormalisation equation}
As seen above the momentum  integral sumrule leads to a specific non integer value
of $\nf$ and as a consequence also for $\beta_0$  and by extension for the
full set of $\beta$ governing the $\alpha_s$ running. It is also
natural that the coupling constant depends on the sum of flavor activities
$l_q$ and not only on flavor number.
%Anyhow $\alpha_s$ and parton evolution are linked by renormalisation group theory.
\subsection{Coefficient functions}
As it is the structure functions and not the parton distributions which
are observable one has to find also a procedure to modify the coefficient
functions.\\
%{\bf remarks}
\begin{itemize}
\item
The transform  
 parton distribution $\rightarrow$ structure function has exactly the same
structure that the one of DGLAP equations, it is obtained by the changes:
$$ \frac{\partial o(Q^2)}{\partial ln(Q^2)} \rightarrow F_o$$
$$ \rp \rightarrow \rc$$
\item
This transform is in fact nothing more than a change of scheme,
an example is going from $\overline{MS}$ to $DIS$ for $F_2$.\\ 
For $F_1$ and $F_3$ the schemes are unnamed but they still exist.
\end{itemize}

From this one may infer that coefficient functions have to be modified
in the same way that splitting functions.\\
In fact it is the importance of kinematic constraints stressed by
one of the R.Thorne papers and its use in the latest schemes which
started this work.\\
{\bf Charged currents}\\
Exactly the same procedure will be used, the phase space only will
change using:
\begin{equation}
l_o=(1+m_o^2Q^{-2})^{-1}
\end{equation}
\section {Theoretical Considerations}
\begin{figure}[here]
\resizebox{1.1\textwidth}{.5\textwidth}
           {\includegraphics{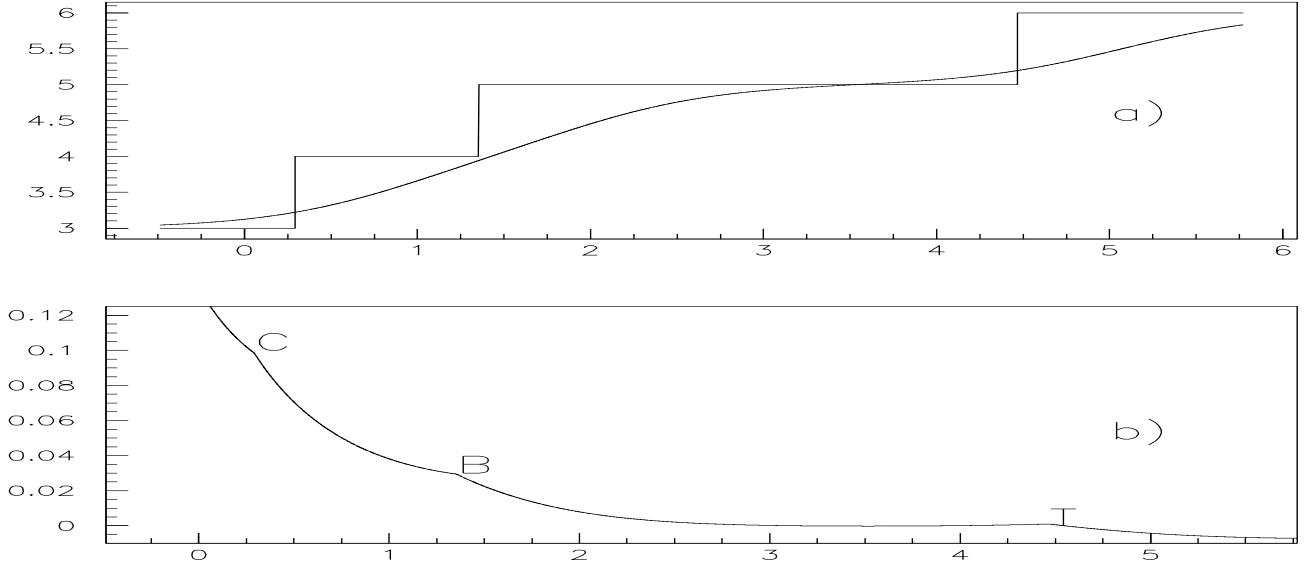}}
\caption{
scheme comparaison with $\alpha_s(M^2_Z)=.118 \qquad \qquad$
a) $\nf$ versus $\log Q^2$ for cfns and massless.
b) Relative difference between  massless and cfns for  $\alpha_s$ versus
$\log Q^2$.
}
\label{as}
\end{figure}

\begin{figure}
 \resizebox{1.1\textwidth}{.5\textwidth}
           {\includegraphics{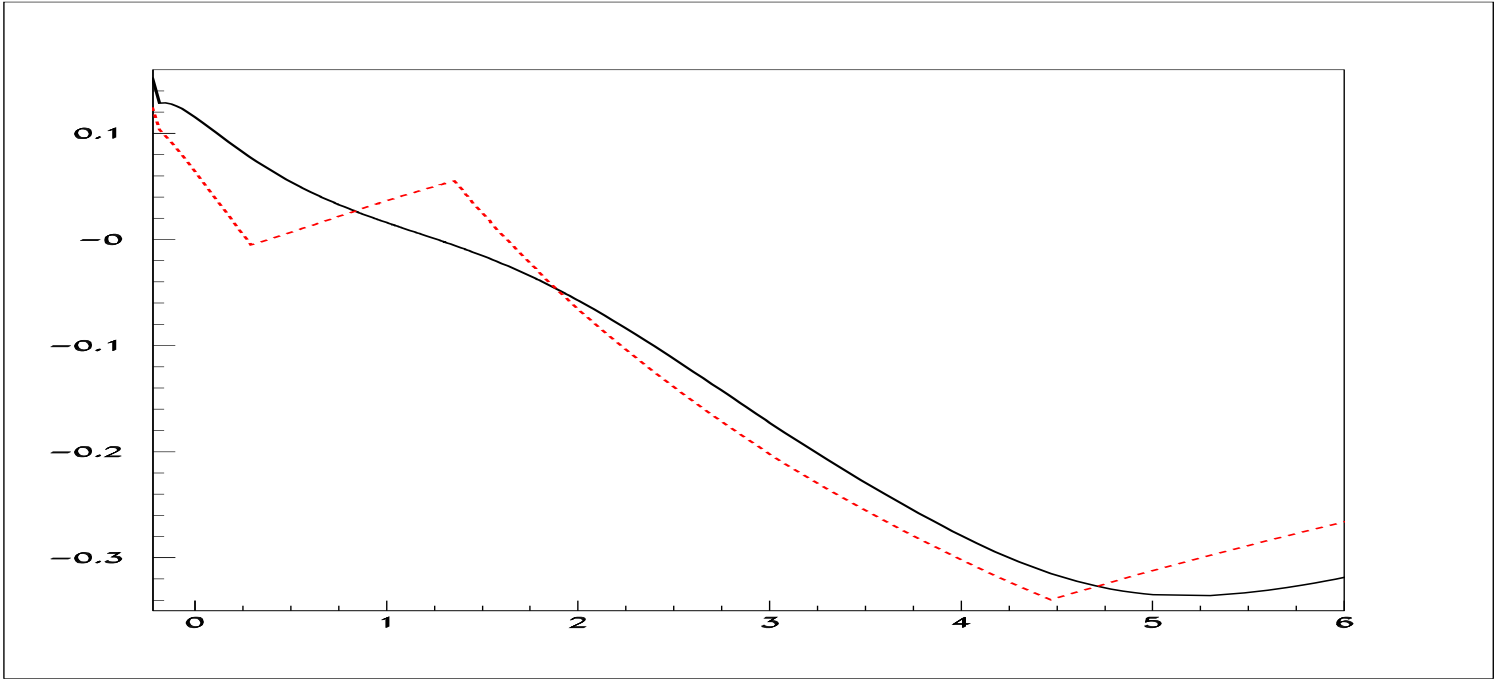}}
\caption{$\rf_3^{e^-p \rightarrow \nu X}$ versus $\log Q^2 \qquad $
cfns: black full, massless: red dotted.}

\label{f2cc}
\end{figure}

Extended  DGLAP  equations have been set up very close to the ordinary ones.
\begin{itemize}
\item
They reduce to it when scale $Q^2$ is far away from 
any heavy quark mass squared.
\item
They are integro differential equations, linear in all the  pdfs which they
use in conjonction with convolution integrals.
\item
They satisfy the kinematic constraints of the heavy partons.
\item
The resulting pdf are continous in $x$ and $Q^2$ at least if the input pdfs at
the initial scale are.
\item
The derivatives of the pdfs are also continous.
\item
The same properties apply for charged and neutral currents structure
 functions $F_1$,$F_2$ and $F_3$ (see Figure \ref{f2cc}). 
\item
When there is only active flavors (all $l_h$ close to 0 or 1)
\hspace{1cm}
$\rs_{oi}(x,Q^2)\rightarrow P_{oi}(x,Q^2,\nf=N_f)$
\item
They are supported by works and concepts which are not really new:
\begin{itemize}
\item
 $\xi$ the scaling variable was used in 
many papers, see \cite{Georgi} as an example.
\item
 \cite{Georgi} use also anomalous dimensions variable with $Q^2$.
Anomalous dimensions leading to splitting functions their arguments
should hold here. They advocate\footnotemark~
\footnotetext{\it they did note have the same name and used it differently
however.} $l_i \approx (1+2m^2_iQ^{-2})^{-1}$.
\item
Moreover  \cite{Georgi} presented a $\beta$-function variable with $Q^2$
(due to  $\beta_0$ ). They advocate here $l_i \approx (1+5m^2_iQ^{-2})^{-1}$.
\item
\cite{Brodsky} have considered $\nf$ to be $Q^2$ and even $\alpha_s$
 order dependent and
 there is still developements on the  $\beta$-function \cite{ddd}.
\item
Small are the differences between massless and continous
behavior for $\alpha_s$ (see Figure \ref{as}).
\item
The procedure leading to satisfaction of the kinematical constraints
as been used lately for coefficient functions in GM-VFNS schemes \cite{Thorne}.
\item
Presented here is the simplest solution to the chosen goal but
it is still possible to make modifications making use of the freedom brought by
the $f_{oi}(Q^2)$s following \cite{Brodsky} example.
\end{itemize}
\end{itemize}

Satisfiying all the points it was designed for, it as some serious advantages:
\begin{itemize}
\item
It does not mix up different $\alpha_s$ orders as do mixed schemes. 
\item
Heavy quarks participate to the evolution when they start to appear,
that is at the beginning and at very low $x$ and even at leading order
in $\alpha_s$.
\item
There is  only internal partons, no externals. 
\item
It covers the charm-bottom region where they are both opening up
(see $\nf$ graph Figure \ref{as}) which is not yet the case in VFNS.
\item
It should give a better treatement of the small x region where heavy quark
pdfs appear first.
\end{itemize}
It seems that ressummation is done, due to evolution, but not
the forward divergence pole subtraction.  
In that case this method should be considered only as a phenomenological
model which may be optimized by adjusting the $f_{oi}$ functions with 
the help of the rates given by \cite{rsn}.
The comparaison between the $F_2^{c \bar c}$ of the two approaches would
however be not that simple, having so different concepts
(quark internal versus external). \\
An important point is that nature has 6 flavors and not 3,4,5 or 6
depending on physicist will, so neglecting that fact is making an 
approximation.
This paper is also making approximations which are certainly valid when 
$\nf(Q^2)$ is close to an integer.
%(at least in a mathematical world where the heavy quark
%masses are well sparated and for any number of flavors)\\
Maybe renormalisation group theory could bring some light to this.
\section {Program implementation}
Modifications to be made to existing programs may be done without changing
their conceptual building, however changes are not that small.
The QCDFIT case (\cite{zp}) will be presented in some details
as an example.\\
{\bf QCDFIT concept}\\
It is a program which works in $x$ space.
It includes an optimizing procedure (Minuit package \cite{minuit}).
it accept a variety of input distributions.
It has a variety of outputs: Pdfs, cross section for lepto production
,Drell-Yang mecanism... 
It precalculate the full evolution saving there a lot of CPU time.
\\
{\bf Pdf representation}\\
QCDFIT uses two grids:
The $x$ grid has linear spacing in $\log(x)$.
There is much more liberty for the $Q^2$ grid. Present version uses 
a $Q^2$ grid approximatly equally spaced in $a_s$. \footnotemark~
\footnotetext{\it A maximum of 20000 $x$ nodes and 600 $Q^2$ nodes has been
 used}
All the functions of $x$ or $Q^2$ or ($x$, $Q^2$) are calculated on these grids.
To know values elsewhere a linear approximation has to be made
(there is specialised routines for that).
One consequence is that distributions are represented by vectors and that
the evolution of the distributions from one $Q^2$ grid value to an other
$Q^2$ grid value is given by matrices.\\
{\bf Kernel representation}\\
They are linear operators applied to the pdfs and so after discretisation have
 a matrix
representation. But due to the linear appoximation used, 
the structure of convolution and more deeply to the
concept of parton branching those matrices are upper triangular band matrices:
their elements are: $\rm_{ij}=m_{i-j}$ with $i \ge j$.\\
As a consequence in QCDFIT they are represented by one dimensional arrays
 and a system of fast and simple routines have been developed to deal with
 kernel multiplication,
inversion, exponentiation, square root and vector (pdf) multiplication.
Notice that target mass correction, higher twist and renormalon a la 
M.Dasgupta and B.R. Webber are  eventually computed in this kernel frame.\\
{\bf QCDFIT Evolution}\\
Integration of the renormalisation group equation is made analytically
for the needed flavors. Integration of $a_s^n$ needed for kernel integration
is made at the same time and with the same technique.\\
For the NS case it is easy to show that the solution of DGLAP involves an
integration of the kernel followed by an exponentiation
(easy and fast operation with QCDFIT concept). For the Singlet case
things are more complicated  and an additional calculation is needed.
\\   
{\bf QCDFIT Modification}\\
The representation of the Pdfs, the kernels, the convolution algebra stay
the same, but the evolution itself has to be modified:
Integration of the renormalisation group equation is now made numerically
as its $\beta$ parameters are functions of $\nf$ and so of $Q^2$.
The transport matrices defined by 
$o(Q^2_{j+1})= \sum_i \tr(o,i)\otimes i(Q^2_j)$
are obtained by integration of the subsystems using:
 \begin{eqnarray}
\tr(o,i) = \prod_{Q^2_j}^{Q^2_{j+1}} \Biggl(
1+ \frac{\partial^2 o}{\partial i \partial ln(Q^2)} \delta ln(Q^2) \Biggr)
\end{eqnarray}
In the product $\delta ln(Q^2)$ has to be small enough to see only 
the rounding errors when increasing the number of $Q^2$ nodes.\footnotemark~
\footnotetext{\it QCDFIT use presently grids from 10000 to 500000 $Q^2$ nodes}

\section{cfns  massless comparaison at NNLO}
\begin{figure}[here]
 \resizebox{1.1\textwidth}{.5\textwidth}
           {\includegraphics{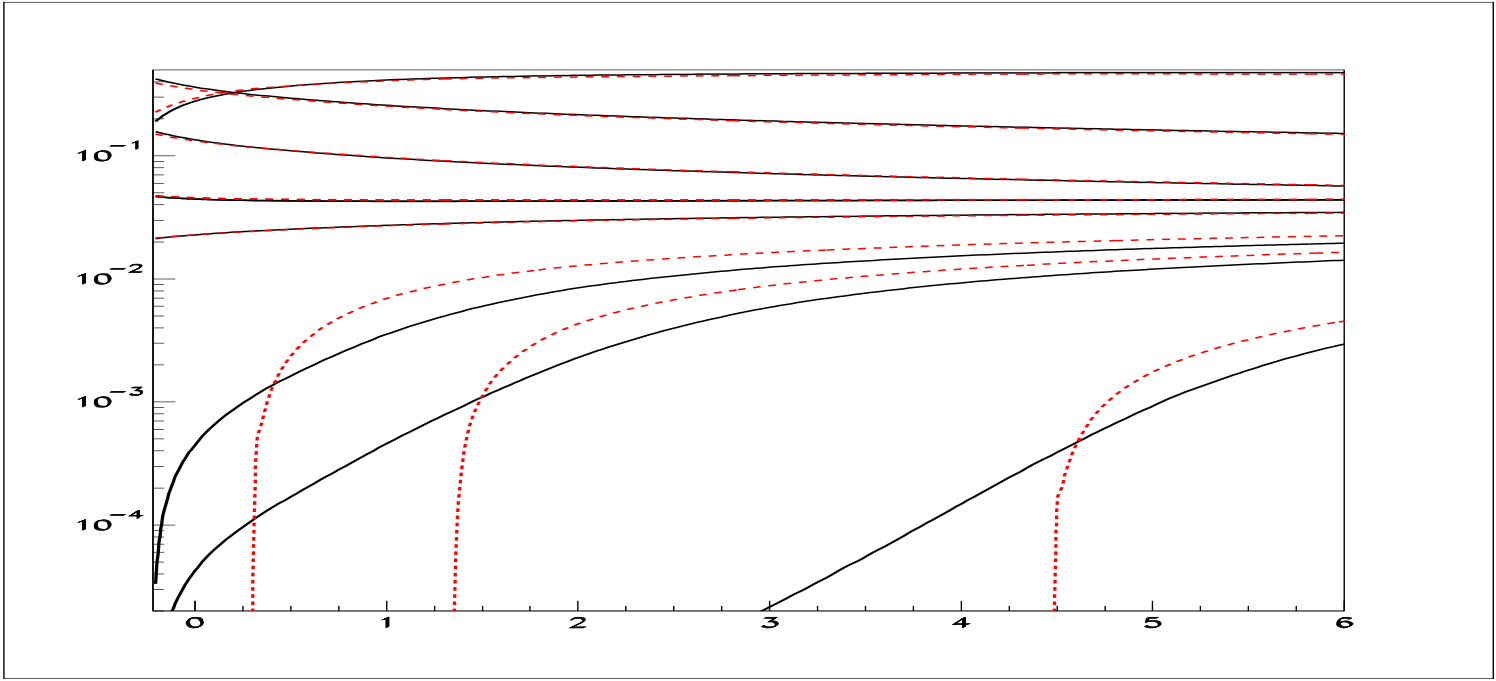}}
\caption{From top to bottom $g,u_{val},d_{val},\bar u =\bar d,s,c,b,t$
momentum fractions versus $\log Q^2. \qquad  $
cfns: black full, massless: red dotted.}
\label{frac}
\end{figure}

\begin{figure}
 \resizebox{1.1\textwidth}{.5\textwidth}
           {\includegraphics{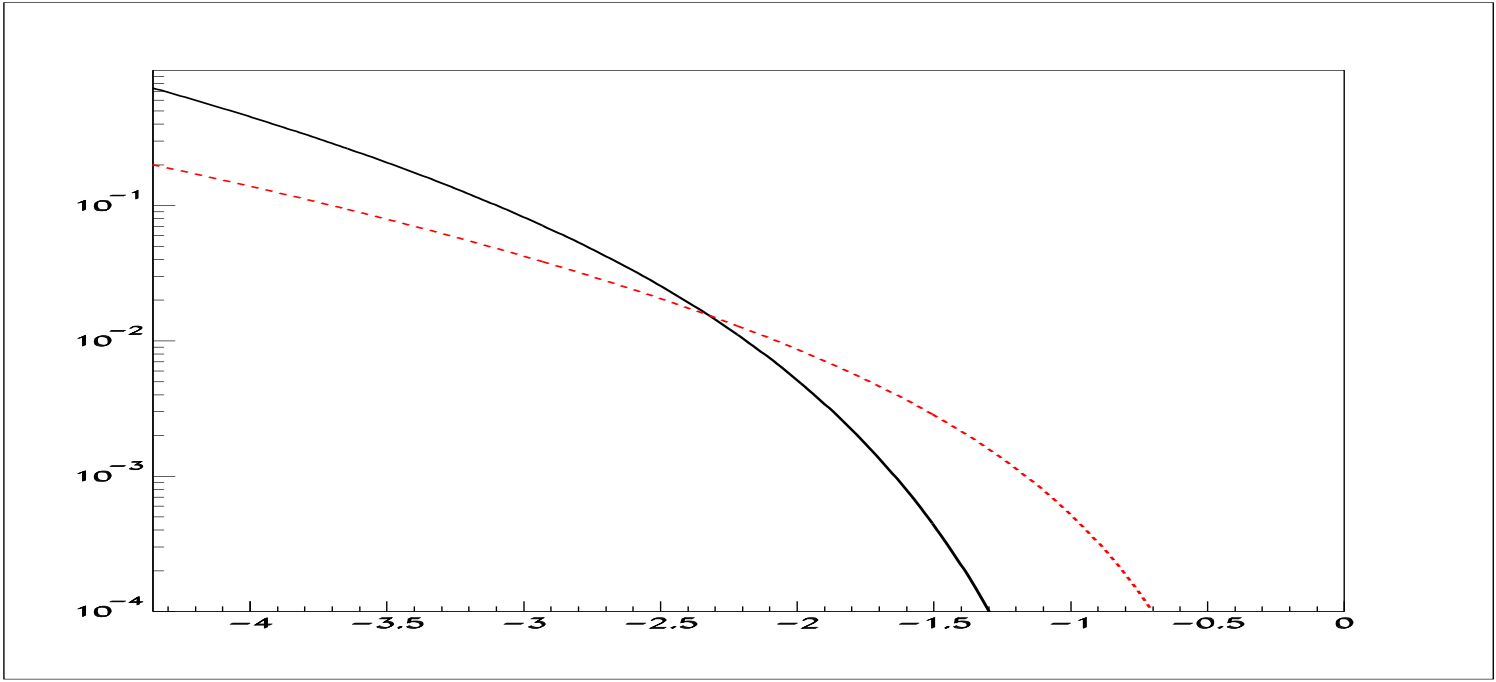}}
 \caption{ Top pdf cfns: black full, massless: red dotted at $Q^2=39800 Gev^2$
where both momentum fractions are equal.}
\label{equal}
\end{figure}

The aim of the exercise is to show what kind of new features might be seen
on cfns and how far they extend away from the transition points of the
 other schemes.\\
{\bf Kinematical range used}\\
Very often the start of evolution $Q^2_{input}$ is chosen just below
the charm masss squared in order to define pdf inputs only for light partons.
But for cfns heavy quarks are always present if the kinematical range permits,
so in order to have only light quarks present , a $Q^2_{input}= 0.6 ~Gev^2$
is used, low enough to justify neglecting all the heavy quarks at input. 
Needless to say that at so low a $Q^2$ the predictive power of the pdfs
is completly absent but it is a parameterless way to get a sensible charm
when out of the non perturbative region.\\
A  data sample made of about 1850 F2 or cross section measurements
extracted from NMC, BCDMS on protons and H1 preliminary is used to fit 
the input pdfs independently for cfns and for massless scheme.
The fitted distributions are $g,u_{val},d_{val}, \bar u = \bar d = \lambda \bar s 
\quad$
with $\int_0^1 x ~dx~ \bar s= 0.53 \int_0^1 x ~dx~ \bar d$ 
at $Q^2 =10  Gev^2$ \cite{Martin:2009iq}\\
{\bf Results}\\
Figure \ref{frac} shows the momentum integral fractions for the
partons species in both schemes as a function of $\log(Q^2)$. Note for
heavy quarks the earlier start and slower rise of cfns. 
Figures \ref{f2cc} shows the cfns  and massless structure function
$\rf_3^{e^-p \rightarrow \nu X}$ at $x=0.01$ versus
 $Q^2$.  The three kinks at $m_h^2$ in massless are due
to the alternative coming in of quarks and antiquarks wih negative sign.
In cfns slope continuity is restored because there in a single DGLAP
system and coefficient functions with no discontinuities.
%Figures \ref{start} shows the cfns heavy quark distributions at $Q^2=m^2_h$
%when the massless ones are still null.
%The somewhat different behavior of charm comes from the rapid change of the
%gluon.
Figures \ref{equal} shows the cfns  and massless top distributions at
$Q^2=39800$ where both momentum fractions are equal,
cfns has a much steeper distribution with  highers values at
small x coming from its early developement and from the kinematical constraint
(not shown is the same effect for charm and beauty).

\section*{Conclusion}
Effects shown are impressive and justify
the idea of evoluting heavy quarks even when they have
to be considered as heavy.\\
Work will continue to assess on more firmer theoretical ground  these ideas.

\end{document}